\documentclass[twocolumn]{aastex631}
\usepackage{listings}
\usepackage{subfigure}
\usepackage{xcolor}
\usepackage{xcolor, soul}
\sethlcolor{yellow}
\usepackage{enumitem}
\hypersetup{backref=true, pagebackref=true, hyperindex=true, colorlinks=true,                
	breaklinks=true, urlcolor= magenta, linkcolor= black, bookmarks=true,                 
	bookmarksopen=false, filecolor=cyan, citecolor=blue, linkbordercolor=blue}
\usepackage{hyperref}

\shorttitle{SSA and Transits}
\shortauthors{Fatheddin and Sajadian}

\usepackage{amsmath}
\begin{document}
\title{Singular Spectrum Analysis of Exoplanetary Transits}

\author[0000-0002-7611-9249]{Hossein Fatheddin}
\affiliation{Leiden Observatory, Leiden University, PO Box 9513, NL-2300 RA Leiden, the Netherlands; \url{Fatheddin@strw.leidenuniv.nl}}
 
\author[0000-0002-0167-3595]{Sedighe Sajadian}
\affiliation{Department of Physics, Isfahan University of Technology, Isfahan 84156-83111, Iran}

\begin{abstract}
Transit photometry is currently the most efficient and sensitive method for detecting extrasolar planets (exoplanets) and a large majority of confirmed exoplanets have been detected with this method. The substantial success of space-based missions such as NASA's Kepler/K2 and Transiting Exoplanet Survey Satellite (TESS) has generated a large and diverse sample of confirmed and candidate exoplanets. Singular Spectrum Analysis (SSA) provides a useful tool for studying photometric time series and exoplanetary transits. SSA is a technique for decomposing a time series into a sum of its main components, where each component is a separate time series that incorporates specific information from the behavior of the initial time series. SSA can be implemented for extracting important information (such as main trends and signals) from the photometry data or reducing the noise factors. The detectability and accurate characterization of an exoplanetary transit signal is principally determined by its signal-to-noise ratio (SNR). Stellar variability of the host star, small planet to star radius ratio, background noises from other sources in the field of observations and instrumental noise can cause lower SNRs and consequently, more complexities or inaccuracies in the modeling of the transit signals, which in turn leads to the inaccurate inference of the astrophysical parameters of the planetary object. Therefore, implementing SSA leads to a more accurate characterization of exoplanetary transits and is also capable of detecting transits with low SNRs ($SNR<10$). In this paper, after discussing the principles and properties of SSA, we investigate its applications for studying photometric transit data and detecting low SNR exoplanet candidates.  

\end{abstract}

\section{Introduction} \label{sec:intro}
Since the first discovery of a Jovian-mass planet around a solar-type star in 1995 \citep{1995Mayor}, the theoretical and observational efforts for detecting and characterizing exoplanets have evolved tremendously. Since exoplanets are extremely faint objects, several indirect methods have been developed for detecting them. So far, the major exoplanet detection methods include \citep{2013Wright, 2023Fatheddin}: Radial Velocity measurements (which is based on observing the Doppler shifts in the spectrum of the planet's host star), Gravitational Microlensing (which is based on the temporal magnification in the light of a source star due to a gravitational binary lens system consisting of a star and a planet) and Transits (which is based on the periodic dimming in the photometric light curve of a star, caused by a passing planet). 

\indent Currently, the transit photometry has been the most efficient method; out of the 5535 confirmed exoplanet detections to date\footnote{\url{https://exoplanetarchive.ipac.caltech.edu/docs/counts_detail.html}}, about 4100 planets were discovered via the transit method. Data from space-based missions, such as Kepler/K2 \citep{2010Borucki, 2014Howell} and the Transiting Exoplanet Survey Satellite (TESS) \citep{2014Ricker}, has provided a large and diverse sample of transiting exoplanets. These missions implement high precision and high cadence time-domain photometric observations of a large sample of stars in our Galactic neighborhood from the stable environment of space \citep{2021Stapelfeldt}.

\noindent For detecting the periodic dimming caused by a transiting exoplanet in the time series photometry of a star, the Box-fitting Least Squares (BLS) algorithm \citep{2002Zucker} is implemented in most cases. In the BLS formalism, the transit signals are modelled with periodic box-shaped functions (upside down boxcar function) with the following parameters: Period, Depth, Duration and Mid-Transit Time (as Reference Time). For determining the best fit of these parameters to the transit signal, a range of periods is considered and the time series is folded to each trial period in that range. The best least squares fit (maximized over depth) for a period determines its corresponding signal power in a resulting periodogram. Finally, the parameters corresponding to the highest power in the periodogram, are considered as the best transit model. 
  
\indent The most important planet detection metric in the transit method is the signal-to-noise ratio (SNR) of the transit signal \citep{2023Kipping}. The SNR is a precise metric based on the properties and characteristics of the transit signal and noise in the photometric observations. In this paper, we estimate the SNR of a transit as the ratio of the measured transit depth ($\delta$) to the averaged noise estimate from the Combined Differential Photometric Precision (CDPP) metric ($\sigma_{cdpp}$) \citep{2000Gaudi, 2018Kunimoto}:
\begin{eqnarray}
\label{eq:SNR}
S/N=\sqrt{N_{tran}}\frac{\delta}{\sigma_{cdpp}}
\end{eqnarray}  
where $N_{tran}$ is the number of transits and $\sqrt{N_{tran}}$ acts as the constant of proportionality. The depth of a transit ($\delta$ in equation \ref{eq:SNR}) is proportional to the ratio of squared planet to star radius \citep{2019Heller} and can be easily measured from the best fit model from the BLS algorithm. 

\indent The main idea behind the Combined Differential Photometric Precision (CDPP) metric is that after removing all long term trends in the light curve, an estimate of the noise can be measured as the averaged fluctuations in data \citep{2011Gilliland}. In order to measure the averaged noise estimate ($\sigma_{cdpp}$), we start by detrending the light curve and removing the outliers. Then, we remove low frequency signals using a Savitzky-Golay filter \citep{1964Savitzky} with window length equal to the duration of the transit. Finally, $\sigma_{cdpp}$ is calculated as the standard deviation of a running mean over the data. The window length of the running mean is also set to the duration of a single transit.

\indent Singular Spectrum Analysis (SSA) is a powerful tool for time series analysis with many applications \citep{Broomhead1986, 1989Vautard, 1992Vautard}. The SSA can be used to decompose  a time series into a sum of its components with separate periodicities occurring on different time scales \citep{2007Hassani}. These components can be grouped together as trend, periodicity and noise signals. The original time series can be reconstructed by summing all of its components.

\noindent The applications of SSA in various fields such as Climate Physics, Economics and Oceanology have been studied extensively \citep[for example, see][]{2015Groth, Coussin2022, 2019Majumder}. But, contrary to its great potential, there are currently very few instances of its use in astronomical data analysis \citep{2016Greco, 2021Weinberg, 2023Thekkeppattu}. 

\indent In this paper, after reviewing the principles and properties of the SSA algorithm in Section \ref{sec:SSA}, we will study its applications to the analysis of the photometric time series data and exoplanetary transits in section \ref{sec:SSA-App}. In Section \ref{sec:LowSNR}, we investigate the detectability and characterization of low SNR transits with implementing the SSA algorithm and report our results. Finally, in the last section, we discuss and summarize the results.

\section{Singular Spectrum Analysis (SSA)} \label{sec:SSA}
The Singular Spectrum Analysis (SSA) is a model-free nonparametric technique for time series analysis with the main assumption that the
time series can be decomposed into a sum of different components such as trend, modulated periodicities and noise \citep{2020Agarwal}. The name of "singular spectrum analysis" refers to the spectrum of eigenvalues in a singular value decomposition (SVD) of the covariance matrix. 

\noindent By decomposing a time series via SSA, one can:
\begin{itemize}
\item Extract the main trends (which can be defined as the long-term change in the level of the time series)
\item Extract the periodicities that form the overall composition of the time series
\item Reconstruct a smoothed version of the time series by denoising the data 
\item Fill in the missing observations or forecast the future trends 
\item Perform a linear frequency filter
\end{itemize}
\indent The first step in performing SSA is embedding the time series into a "Trajectory matrix" (which is explained in section \ref{subsec: Traj}). Then we can decompose that trajectory matrix into a collection of eigenvalues and eigenvectors with the SVD \citep{Golyandina2013}. 

\noindent The SSA algorithms for decomposing and reconstructing the time series based on its trajectory matrix are explained in sections  \ref{subsec: Dec} and \ref{subsec: Rec}, respectively.

\subsection{The Trajectory Matrix} \label{subsec: Traj}
For performing SSA, we need a procedure that takes a univariate time series and turn it into a multivariate set of observations \citep{1996Elsner}.

\noindent We start by considering a time series of length $N$ as $P_N=(p_1,p_2,...,p_N)$. We also consider a fixed integer $L$ as the "window length", where: $2<L<N/2$. Then we form column vectors (each with the length L) from subsets of the time series as:
\begin{eqnarray}
\label{eq:CVec}
\textbf{X}_i = (p_i,..., p_{i+L-1})^T
\end{eqnarray} 
where $i = 1, 2,..., K $ and $K = N-L+1$. 

\noindent Now, we can use these vectors to construct the $L \times K$ dimensional trajectory matrix:
\begin{eqnarray}
\label{eq:Traj}
\textbf{X}=[\textbf{X}_1:...:\textbf{X}_K]=
\begin{bmatrix}
p_1 & p_2 & p_3 & \dots & p_K\\
p_2 & p_3 & p_4 & \dots & p_{K+1}\\
p_3 & p_4 & p_5 & \dots & p_{K+2}\\
\vdots & \vdots & \vdots & \ddots & \vdots\\
p_L & p_{L+1} & p_{L+2} & \dots & p_{N}
\end{bmatrix}_{L \times K}
\end{eqnarray}
We notice that this matrix has equal anti-diagonal arrays which is a characteristic of the Hankel matrices \citep{1983Aoki}.

\indent In the SSA formalism, it is necessary that the window length ($L$) is sufficiently large, so that each $\textbf{X}_i$ column vector (equation \ref{eq:CVec}) would incorporate an essential part of the initial time series and we can separate underlying periodicities from the main trends. If $L$ is large enough, we are able to consider each $\textbf{X}_i$ vector as a separate time series which embodies the dynamics of certain characteristics from the original time series ($P_N$).

\indent It is interesting to note that this notion of embedding and constructing the trajectory matrix is also very useful in the context of data analysis for the non-linear dynamical systems and chaos theory \citep{1980Packard, 1981Takens}. In this context, the variables that describe the evolution of the dynamical system (which are a set of $N$ first-order differential equations) are reduced to a single differential equation of $N^{th}$ order by successive differentiation. So, a difficult multi-variable description of the system is replaced with a single-variable description \citep{1996Elsner}.

\subsection{Decomposition} \label{subsec: Dec}
At this point, we can decompose the trajectory matrix $\textbf{X}$ (equation \ref{eq:Traj}) with the SVD algorithm. In the SVD formalism \citep{1980Klema}, factorization of the matrix $\textbf{X}$ is written as:
\begin{eqnarray}
\label{eq:SVD}
\textbf{X}= \textbf{U}\mathbf{\Sigma}\textbf{V}^T
\end{eqnarray} 
\noindent  where \textbf{U} and \textbf{V} (referred to as "left" and "right" singular vectors) are orthogonal unitary matrices with dimensions $L\times L$ and $K\times K$, respectively. $\textbf{V}^T$ is the transpose matrix of $\textbf{V}$ and $\mathbf{\Sigma}$ is a $L\times K$ rectangular diagonal matrix.

\noindent Furthermore, we can use equation \ref{eq:SVD} and some basic linear algebra to show the following relations:
\begin{eqnarray}
\label{eq:U}
\textbf{X}\textbf{X}^T= \textbf{U}\mathbf{\Sigma}\textbf{V}^T\textbf{V}\mathbf{\Sigma}^T\textbf{U}^T=\textbf{U}(\mathbf{\Sigma}\mathbf{\Sigma}^T)\textbf{U}^T
\\
\label{eq:V}
\textbf{X}^T\textbf{X}= \textbf{V}\mathbf{\Sigma}^T\textbf{U}^T\textbf{U}\mathbf{\Sigma}\textbf{V}^T=\textbf{V}(\mathbf{\Sigma}^T\mathbf{\Sigma})\textbf{V}^T
\end{eqnarray}  
From equation \ref{eq:U}, one can simply show that the columns of the $\textbf{U}$ are eigenvectors of $\textbf{X}\textbf{X}^T$ and that they form an orthonormal basis set spanning the subsets of the time series in the columns of the trajectory matrix.

\noindent Similarly, from equation \ref{eq:V}, we can show that the columns of $\textbf{V}$ are eigenvectors of $\textbf{X}^T\textbf{X}$ and that they form an orthonormal basis set spanning the subsets of the time series in the rows of the trajectory matrix.

\noindent Additionally, from both equations \ref{eq:U} and \ref{eq:V}, we see that the elements of $\mathbf{\Sigma}$ (referred to as singular values) are the square roots of the non-zero eigenvalues of $\textbf{X}\textbf{X}^T$ or $\textbf{X}^T\textbf{X}$.

\indent The trajectory space of the time series (defined by $\textbf{U}$ and $\textbf{V}$) is at most L-dimensional. This occurs when all of the columns in the trajectory matrix are linearly independent. But, if there are any columns that are linearly dependent, we would have arrays with zero values in the singular values matrix ($\mathbf{\Sigma}$) which consequently means that the dimension of the trajectory space would be less than $L$. To avoid this, we only consider the trajectory space as d-dimensional, where d is the rank of the trajectory matrix. The rank of a matrix is defined as the maximal number of linearly independent columns of that matrix \citep{2014Banerjee}. 

\indent Finally, the decomposition of the trajectory matrix can be written as:
\begin{eqnarray}
\label{eq:deco}
\textbf{X}= \textbf{X}_1+...+\textbf{X}_d = \sum^d_{i=1}\sqrt{\lambda_i}U_iV_i^T
\end{eqnarray} 
where $(\lambda_1,\lambda_2,...,\lambda_d)$ are the eigenvalues of $\textbf{X}\textbf{X}^T$ or $\textbf{X}^T\textbf{X}$. In the SSA formalism, $X_i= \sqrt{\lambda_i}U_iV_i^T$ are called "elementary matrices" and $(\sqrt{\lambda_i},U_i,V_i^T)$ are called "Eigen-triples".

\noindent It is also worth noting that $\sqrt{\lambda_i}$ can be interpreted as a scaling factor that indicates the relative importance of each eigentriple in the expansion of $\textbf{X}$ (for more information, see: \cite{2001Golyandina}).

\subsection{Reconstruction} \label{subsec: Rec}
Now that we have decomposed the initial time series ($P_N$) into its elementary matrices (equation \ref{eq:deco}), we need a procedure that enables us to reproduce $P_N$ as the sum of $m$ reconstructed time series:
\begin{eqnarray}
\label{eq:RTS}
P_N= \sum^m_{k=1}\tilde p^{(k)}_n 
\end{eqnarray}   
where $\tilde p_n^{(k)}$ are separate component time series from $P_N$.

\noindent We start by partitioning the indices in equation \ref{eq:deco} $(1,...,d)$ into $m$ disjoint subsets $I_1,...,I_m$. If $I=(i_1,...,i_p)$, then the corresponding $\textbf{X}_I$ matrix is defined as:
\begin{eqnarray}
\label{eq:COP}
\textbf{X}_I= \textbf{X}_{i1}+...+\textbf{X}_{ip}
\end{eqnarray}
If we compute the resulting matrices $\textbf{X}_I$ for each $I \in (I_1,...,I_m)$, the grouped expansion of $\textbf{X}$ can be written as:
\begin{eqnarray}
\label{eq:COI}
\textbf{X}= \textbf{X}_{I1}+...+\textbf{X}_{Im}
\end{eqnarray}

Since our intention is to reconstruct the component time series with length $N$ ($\tilde p_n^{(k)}$) from Each matrix $\textbf{X}_{Ij}$, we implement a method called "Diagonal averaging" \citep{2007Hassani}. In diagonal averaging, the values of the reconstructed time series $\tilde p_n^{(k)}$ are defined as averages of the corresponding anti-diagonals of the matrices $\textbf{X}_{Ij}$:
\begin{eqnarray}
\label{eq:COI}
  A_{k} =
\begin{cases}
      \frac{1}{k}\sum^k_{m=1} X_{m,k-m+1} & \text{for $1 \le k<L$}\\
      \frac{1}{L}\sum^L_{m=1} X_{m,k-m+1} & \text{for $L \le k\le K $}\\
      \frac{1}{N-k+1}\sum^{N-K+1}_{m=k-K+1} X_{m,k-m+1} & \text{for $K< k\le N$}
\end{cases} 
\end{eqnarray}
where $A_{k}$ are the elements of time series $\tilde p_n^{(k)}$. The resulting component time series from diagonal averaging are the main and final result of the SSA algorithm and by adding them together, we get the original time series (equation \ref{eq:RTS}).

\indent In order to understand the relationship between each component time series and whether if they should be grouped together or separately (as Trend, Periodicities or Noise), we use the "W-Correlation Matrix" \citep{2017Bgalo}. Let us consider two component time series $\tilde p_i$ and $\tilde p_j$ derived from matrices $\tilde X_{Ii}$ and $\tilde X_{Ij}$ respectively. The weighted inner product for these two time series is:
 \begin{eqnarray}
\label{eq:Wp}
(\tilde p_i,\tilde p_j)_w= \sum_{k=1}^N w_k \tilde p_{i,k} \tilde p_{j,k}
\end{eqnarray}   
where $w_k$ are the weights, defined as:
\begin{eqnarray}
\label{eq:wk}
  w_{k} =
\begin{cases}
      k & \text{for $1 \le k<L$}\\
      L & \text{for $L \le k\le K $}\\
      N-k+1 & \text{for $K< k\le N$}
\end{cases} 
\end{eqnarray} 
Now, we can define the $L\times L$ W-correlation matrix ($W_{corr}$) for every component time series. The arrays of this matrix are:
 \begin{eqnarray}
\label{eq:Wcorr}
W_{i,j}=\frac{(\tilde p_i,\tilde p_j)_w}{||\tilde p_i||_w||\tilde p_j||_w}
\end{eqnarray} 
\\ From this equation, it can be easily shown that if $W_{ij}=0$, the two component time series are orthogonal and completely separable (i.e. should not be grouped together). Consequently, if $W_{ij}$ is close to 1, the two components are correlated and should be grouped together. 
 
\section{SSA of Photometric Time Series} \label{sec:SSA-App} 
Until now, we have discussed and reviewed the SSA algorithm for an arbitrary time series. A very important category of time series in astronomy, is the time-domain photometry; which is the study of how the light (in numerous passbands or filters) from astronomical objects change over time.

\noindent The variations in a photometric time series (or light curve) of an star can reveal many useful information, including but not limited to: periodic and semi-regular stellar pulsations \citep{1974cox}, stars with outbursts \citep{2010Forgan}, asteroseismology studies of variable stars \citep{2016Mauro}, eclipses and occultation (in binary stars or planetary transits) \citep{2006Willems} and gravitational microlensing \citep{2012Mao, 2023Sajadian}.

\indent  After the immense success of NASA's Kepler/K2 space-based missions, TESS is providing a large and diverse amount of photometric time series. Unlike Kepler/K2 which covered a fixed region of sky \citep{2010Borucki}, TESS is continuously surveying $85\%$ of the sky divided into sectors. Each sector is observed for two orbits of the telescope around the Earth (about 27 days on average) with 2 seconds cadence \citep{2014Ricker}. 

\noindent In this section, we investigate the implementation and applications of the SSA algorithm (section \ref{sec:SSA}) for simulated photometric time series and an actual light curve from TESS (TIC $94727308$) as our case studies.

\subsection{Case Study 1: Simulated Photometric Time Series} \label{subsec: CS1}

 \begin{figure*}[]
 \center
	\subfigure[]{\includegraphics[angle=0,width=1.0\textwidth,clip=]{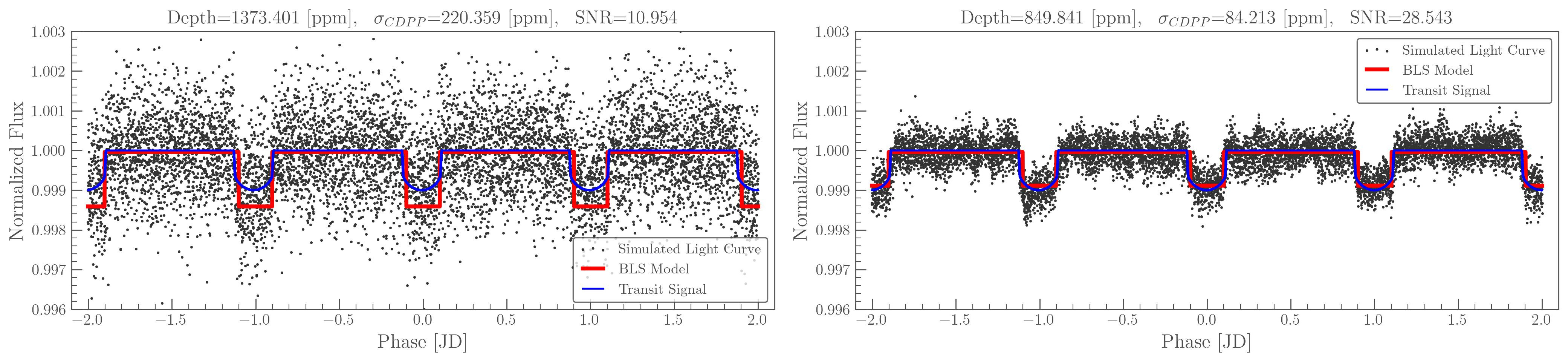}\label{fig:fig17}}\\		
	\subfigure[]{\includegraphics[angle=0,width=1.0\textwidth,clip=]{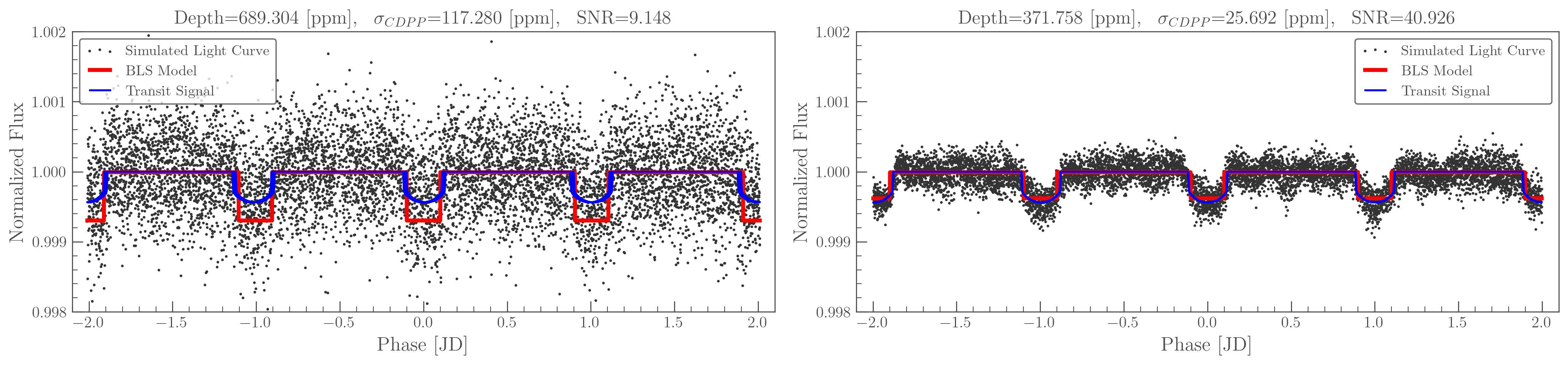}\label{fig:fig18}}\\		
	\subfigure[]{\includegraphics[angle=0,width=1.0\textwidth,clip=]{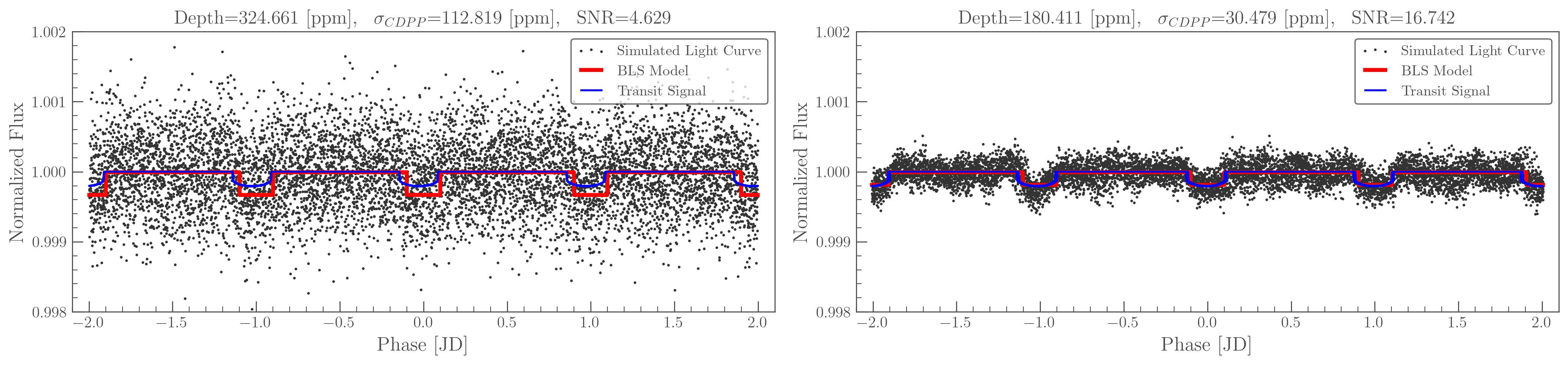}\label{fig:fig19}}	
	\caption{Left Panels: Simulated light curves of low SNR planetary transits with different depths ($\delta$) and noise levels ($\sigma_{cdpp}$). In each simulation, the period is set to one day. The red curve in each figure corresponds to the best transit model derived from the BLS algorithm. The blue curves correspond to the actual transits without the noise factors. Right Panels: The resulting light curves after the SSA de-noising. Before SSA, differences can be noted between the BLS models and the actual transits. These differences are resolved after SSA. On top of each figure, Depth, $\sigma_{cdpp}$ and SNR of the transits are reported.     \label{fig:slc}}
\end{figure*}

In left panels of figure \ref{fig:slc}, we have generated three artificial light curves for stars with different exoplanetary transit signals. Each light curve contains a noise-like background feature which corresponds to the variability of the host star and the photometric measurement error of the telescope. The periods are all set equal to one day, but we have considered a different planetary radius in each simulation in order to simulate different transit depths.

\noindent As we can see in each light curve in the left panel, there is a discrepancy between the actual transit signals (blue curves) and the BLS transit model (red curves) derived from the "noisy" data. Specifically, simulations \ref{fig:fig17} and \ref{fig:fig18} show considerable differences between actual transit depths and depths derived from the model. As it was mentioned in the introduction, the depth of a transit is directly used for inferring the planetary radius, hence even a slight discrepancy between the model and actual transit depth may yield inaccurate results for the astrophysical characteristics of the planet. For simulation \ref{fig:fig19}, where the signal has a very low SNR (4.63), there is also a difference in mid-transit reference times of the model and signal.

 \begin{figure}[]
	\subfigure[]{\includegraphics[angle=0,width=0.45\textwidth,clip=]{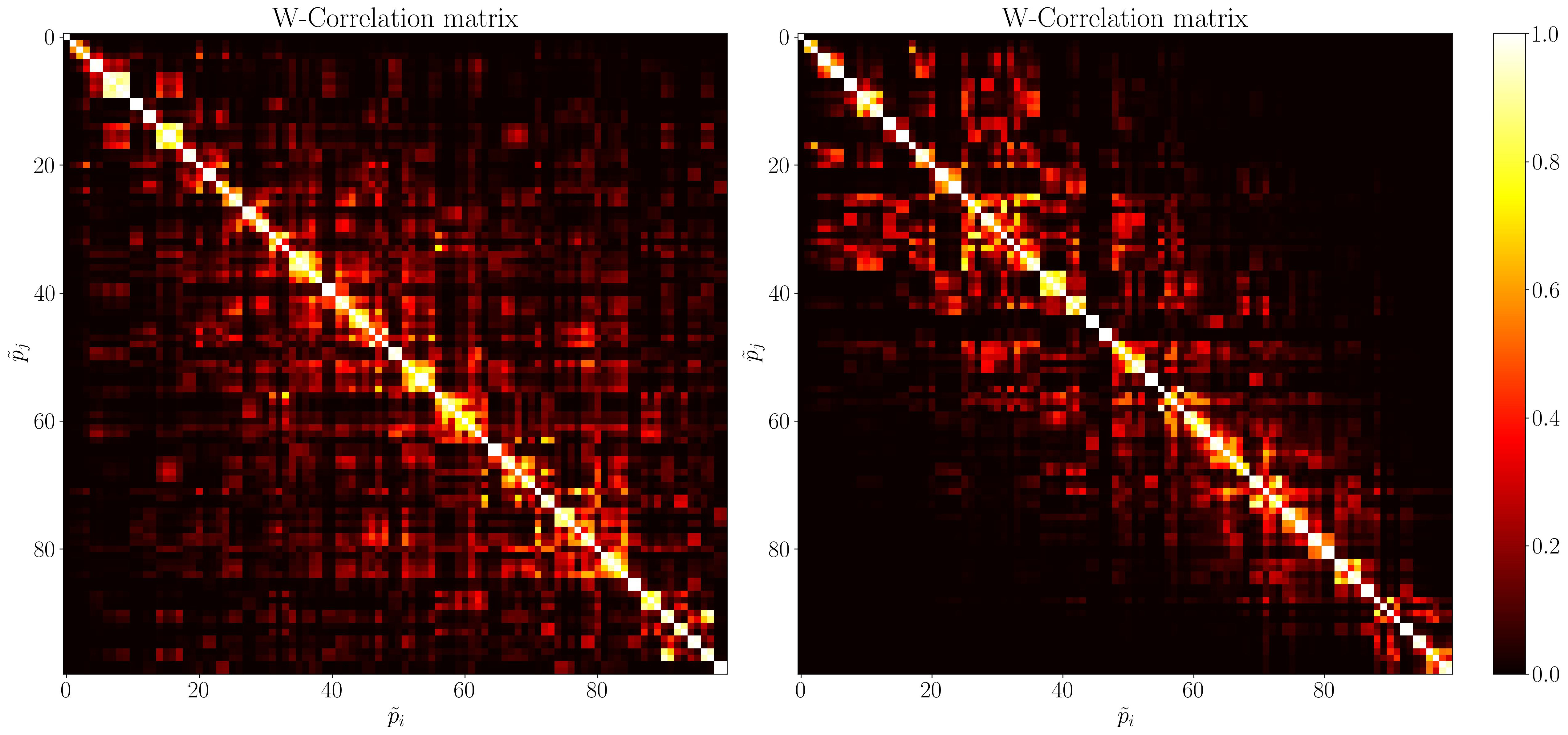}\label{fig:fig20}}\\		
	\subfigure[]{\includegraphics[angle=0,width=0.45\textwidth,clip=]{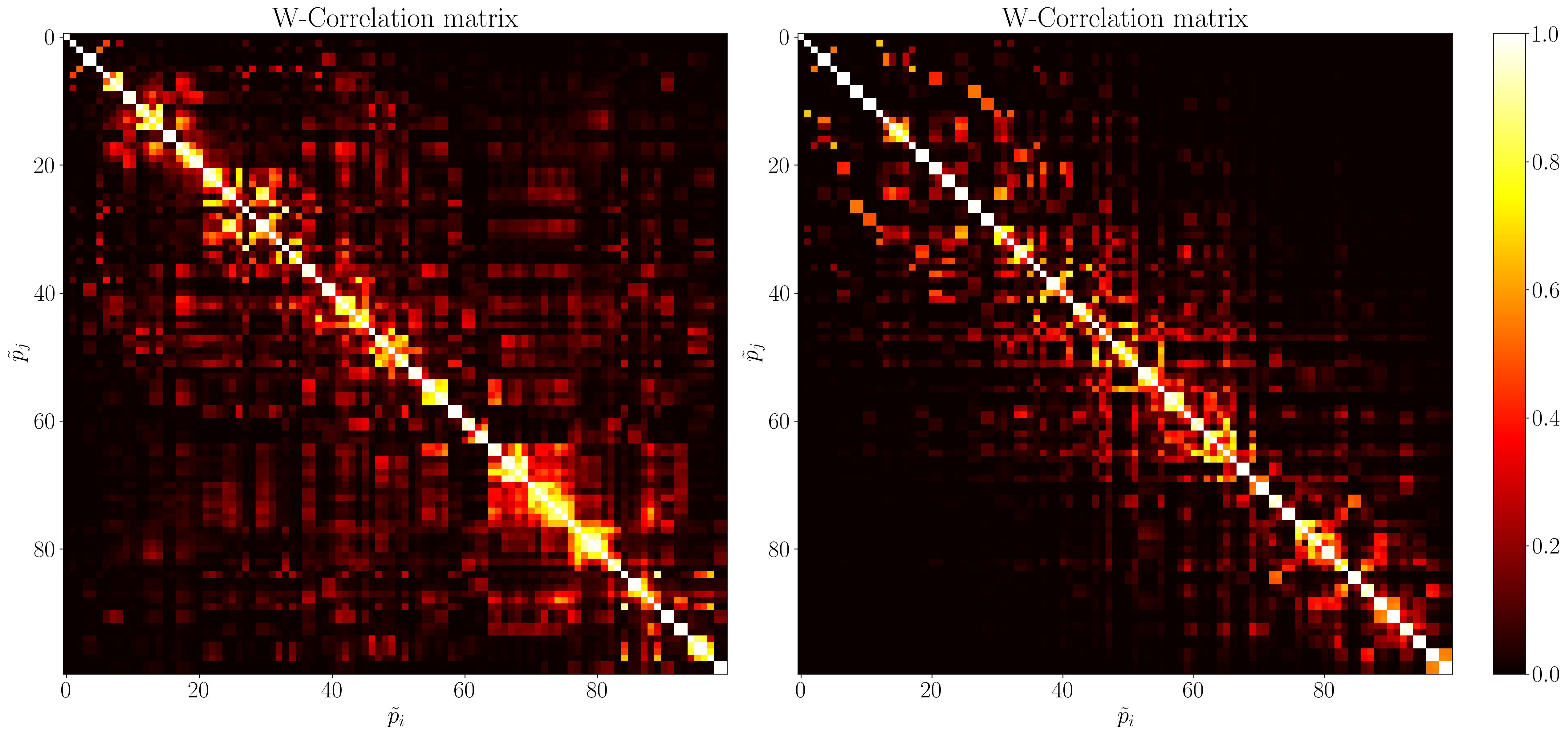}\label{fig:fig21}}\\		
	\subfigure[]{\includegraphics[angle=0,width=0.45\textwidth,clip=]{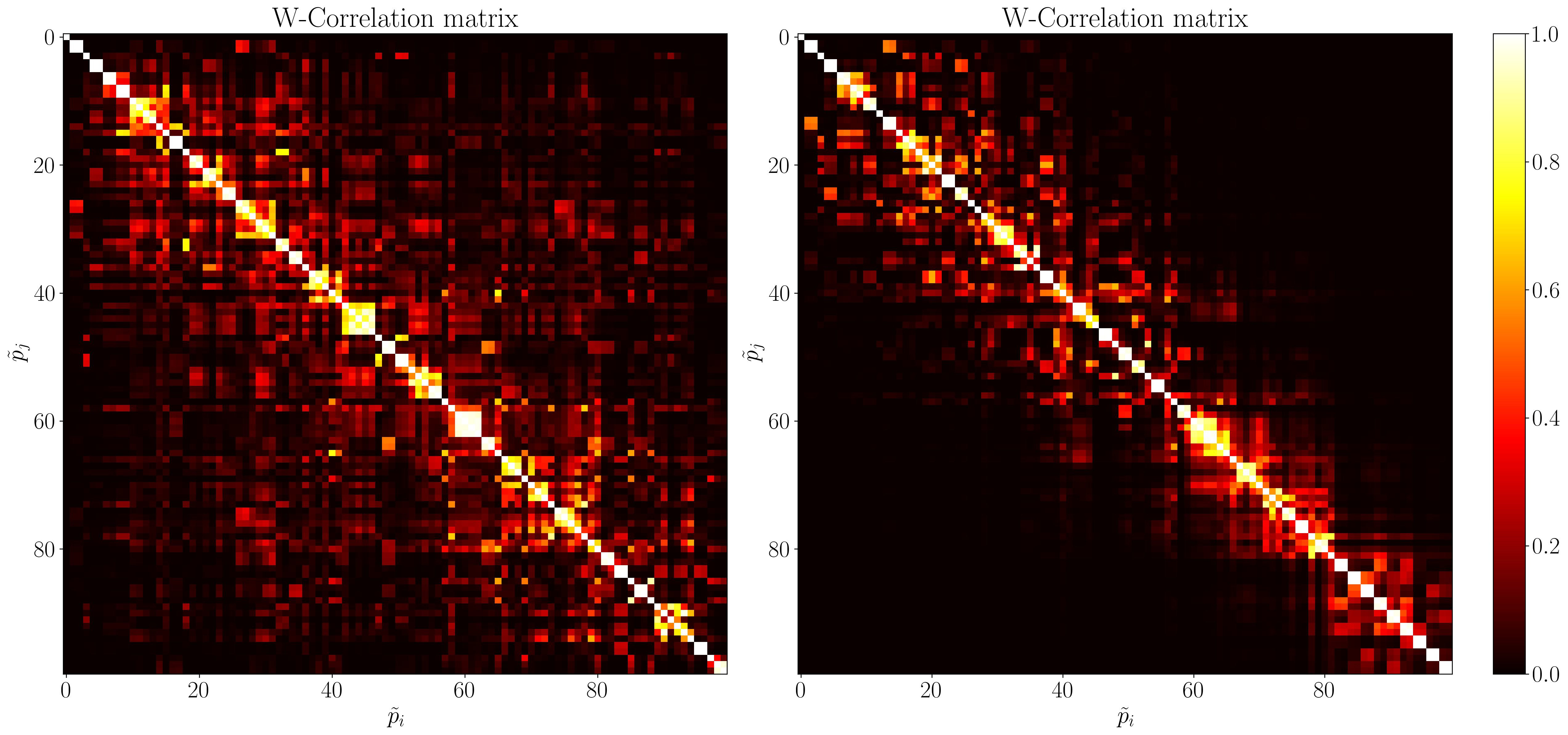}\label{fig:fig22}}	
	\caption{Left Panels: Visualizations of W-correlation matrices for SSA components of the simulations. It can be noted that some component time series show weak correlations with most of the other components which represents their noise-like behaviour. Right Panels: W-correlation matrices of the de-noised light curves. The number of noise components have decreased after SSA. The colorbar at the right of each figure shows the mapping from scalar weighted correlations in the W-correlation matrix to colors. \label{fig:swc}}
\end{figure}

\indent For each simulation, we perform SSA with window length $L=100$. We note that since the length of each time series is $N=8000$, our choice of L is well below N/2. The resulting $100\times 100$ W-correlation matrix visualizations for $100$ component time series of each simulation are shown in the left panels of figure \ref{fig:swc}. In these visualizations, we see that some components are slightly correlated with most of the other component time series. This can be explained by considering the fact that the noise factors have no discernible "separable" structures (regardless of how large the window length is) and as a result, they have a small non-zero weighted inner products with most of the other components.
 
\noindent In order to separate the noise components, we loop over the W-correlation matrix while counting the number of weekly correlated ($W_{ij}<0.2$) components for each component. If this number is larger than a critical value, we omit the corresponding component. The idea behind this procedure is that we are keeping the components that were either completely separated from the other components or have strong correlations, while deleting the components with mostly negligible correlations (indication of their noise-like behaviour). Finally, we can group the remaining components to reconstruct a de-noised version of the time series (section \ref{subsec: Rec}). 

\indent For each simulated light curve, the de-noised version is illustrated in the right panels of figure \ref{fig:slc}. As we can see, after performing the de-noising procedure, we have less fluctuations (lower $\sigma_{cdpp}$) in the light curves and the new BLS transit models, which are derived from the de-noised data, match the real transit signals. In each case, the SNR (equation \ref{eq:SNR}) has also increased substantially. In particular, the transit signal from simulation \ref{fig:fig18} was undetectable before the SSA, as the SNR of its signal was well below the detection criteria ($SNR>10$) for a transiting exoplanet.  This implies that SSA can be implemented to detect low SNR exoplanetary signals in photometric time series (which will be discussed in section \ref{sec:LowSNR} ).

\indent For comparison, we can now perform SSA on the de-noised versions of the simulated data and calculate the corresponding W-correlation matrices (right panels of figure \ref{fig:swc}). As we can see in the new visualizations, the number of components with noise-like behaviours has decreased, indicating a de-noised version of the original time series.
 
\subsection{Case Study 2: TIC 94727308} 
 \begin{figure*}[]
	\subfigure[]{\includegraphics[angle=0,width=0.49\textwidth,clip=]{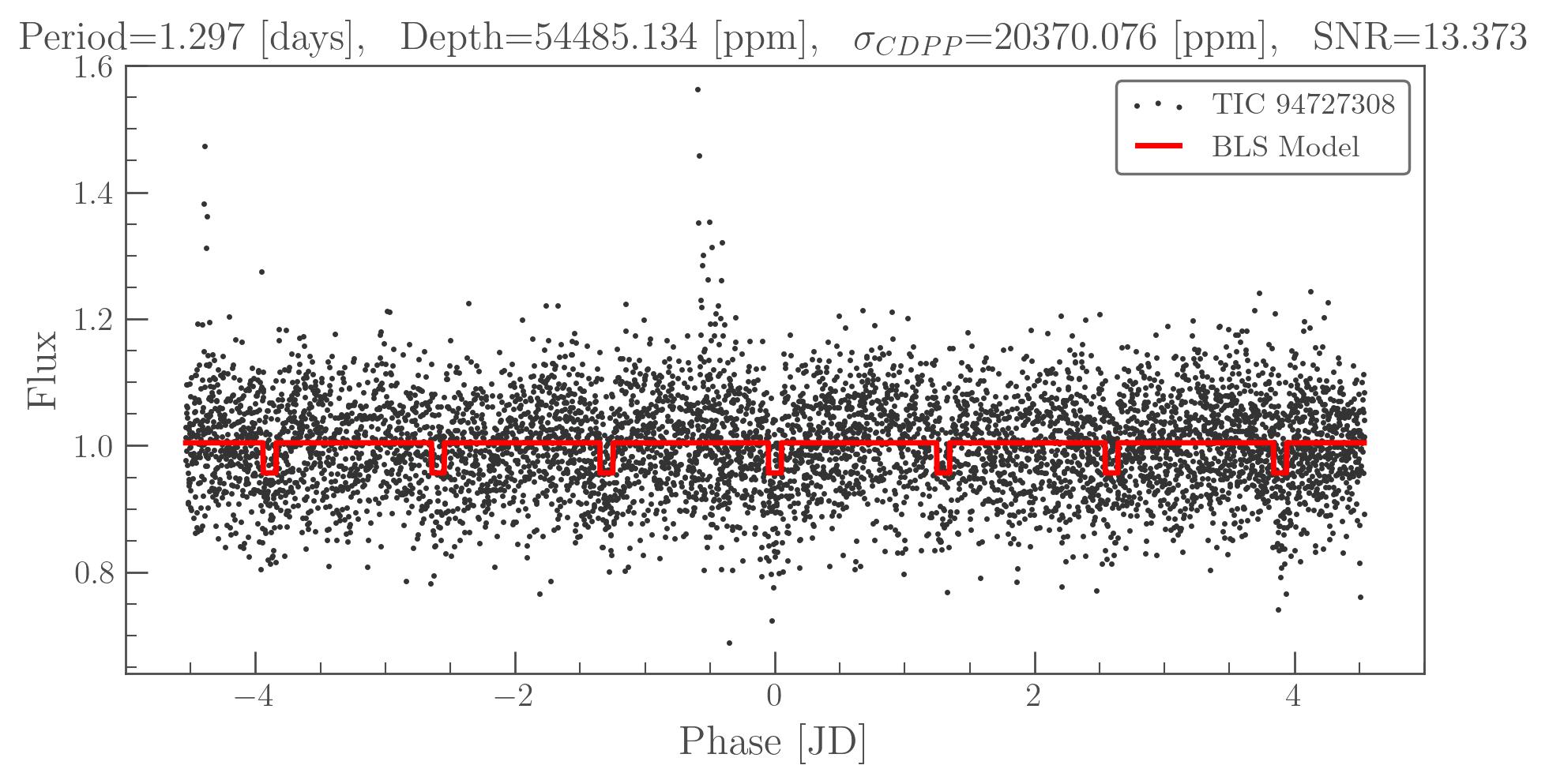}\label{fig:fig6}}		
	\subfigure[]{\includegraphics[angle=0,width=0.49\textwidth,clip=]{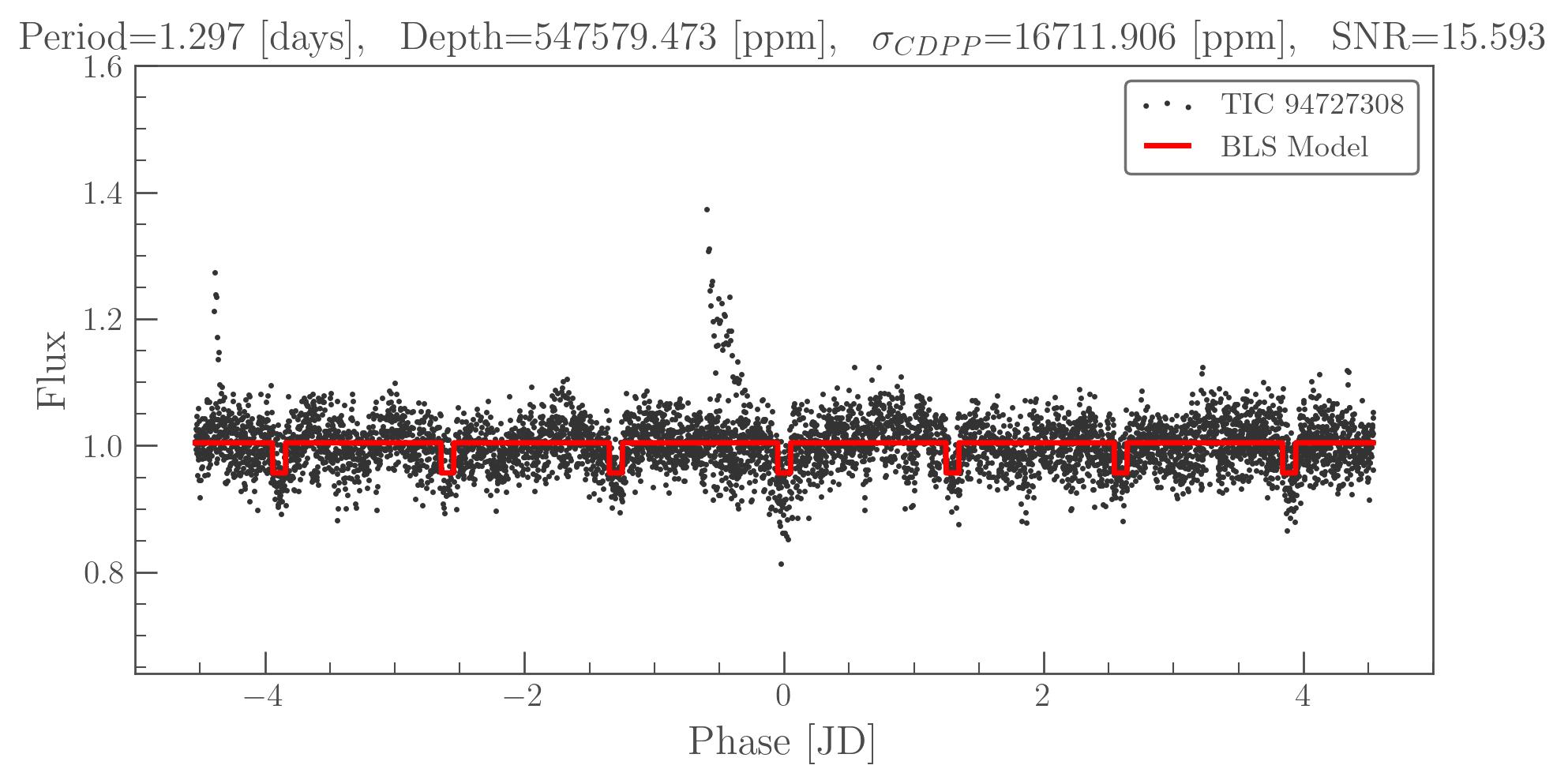}\label{fig:fig7}}		
	\subfigure[]{\includegraphics[angle=0,width=0.52\textwidth,clip=]{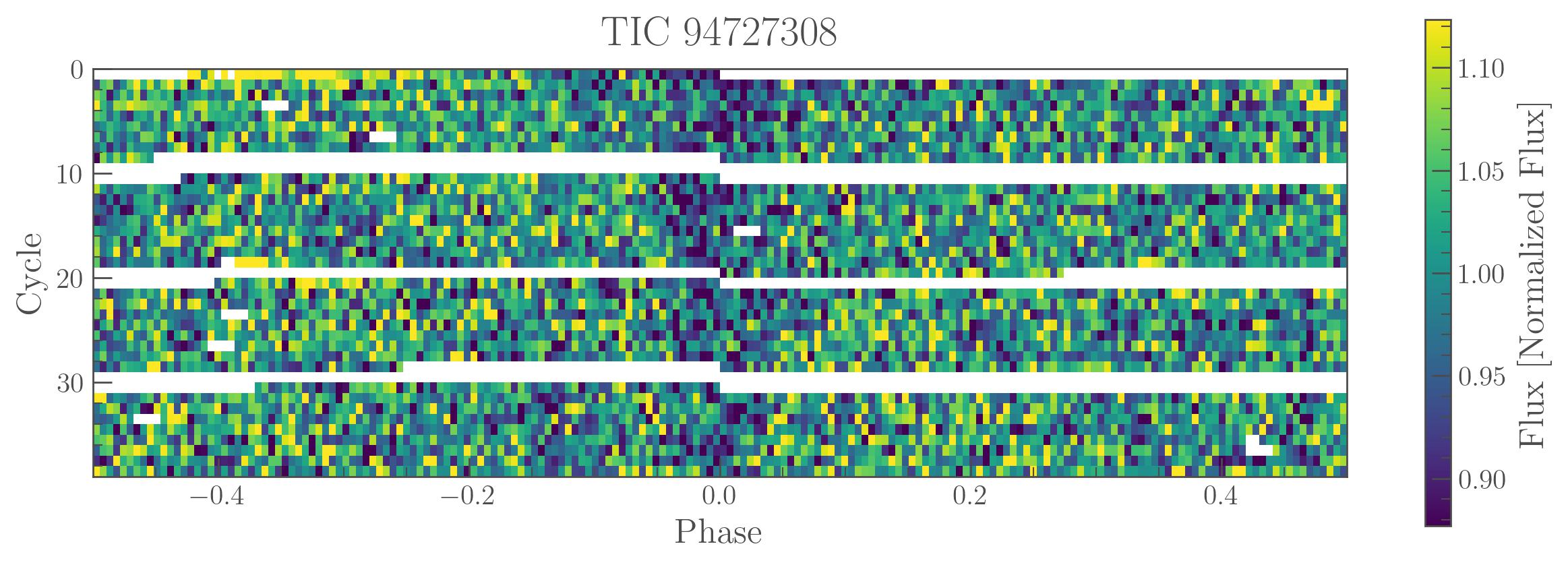}\label{fig:fig8}}	
	\subfigure[]{\includegraphics[angle=0,width=0.52\textwidth,clip=]{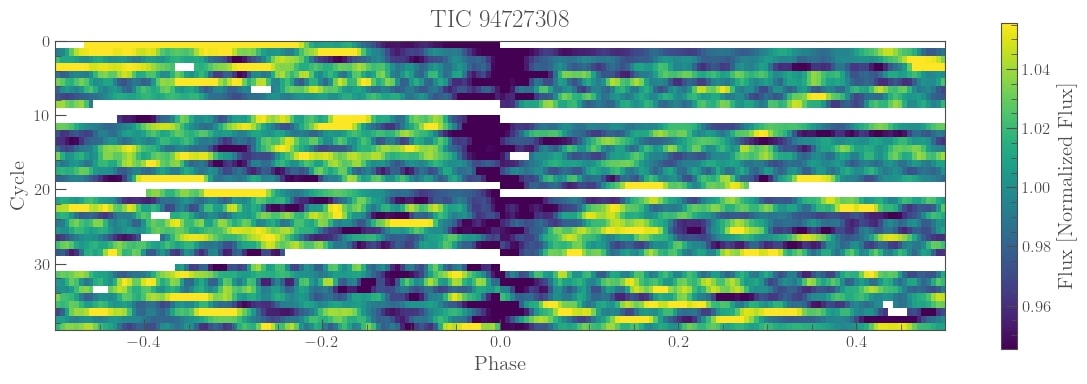}\label{fig:fig9}}		
	\caption{Figure (a) and (c) show the initial light curve and the river plot for TOI-$4622$, respectively. The light curve is folded over seven transits and the river plot shows the evolution of the noisy periodic transit signals over the course of the entire light curve. Figure (b) and (d) show the denoised versions of these plots after the implementation of SSA. The SNR of the transit has increased from 13.37 to 15.59.  \label{fig:lc2}}
\end{figure*}  

\indent TIC $94727308$ is the photometric time series for the TESS Object of Interest (TOI) $4622$ (Gaia DR3 $170033017904053376$). TOI-$4622$ is a very faint star with  I-band apparent magnitude of $16.1779 \pm 0.009$ (from TESS) and G-band mean magnitude of $17.621546$ (from Gaia). Additionally, data from Astrophysical parameters table of the Gaia third data release \citep{Gaia2023} indicate that this star is a faint young M dwarf.

\noindent The TESS photometry data for TOI-$4622$ includes two $600$ seconds exposure time observations from sectors $43$ and $44$ conducted in $2021$. After normalizing and stitching (via the \textsc{LightKurve} package) these time series, the BLS periodogram reveals a relatively low SNR transit signal with period $P=1.297$ days. We fold the detrended light curve over a time interval of $7\times P$ in order to inspect and compare seven transits in figure \ref{fig:fig6}. As we can see, the transit signals seem very weak and they almost resemble noise rather than signal. 

\noindent For an additional and more insightful visualization, we also make the "river plot" for this time series (figure \ref{fig:fig8}). River plots depict the relative fluctuations in flux values by splitting the light curve into parts of equal length (i.e. the transit period), and displaying them side by side \citep{2018Lightkurve}. This allows us to compare these variations over the course of an entire light curve. The river plot in figure \ref{fig:fig8} shows a noisy periodic dimming in the light of the TOI-$4622$.

\indent As described in the previous case study, We perform SSA decomposition on the time series in order to reconstruct a de-noised version of the light curve, by excluding the noise-like components.

\noindent After reconstructing the de-noised version of the time series, we once again fold the data and make the river plot (figures \ref{fig:fig7} and \ref{fig:fig9} respectively). These figures indicate that by omitting the noise components in SSA, we have successfully retrieved the exoplanetary transit signal which was not accurately detectable due to the faintness of the host star. We discuss the further implementation of this de-noising algorithm for detecting and modeling low SNR exoplanets in the following section.   

\section{Low SNR Candidates from TESS} \label{sec:LowSNR}
 \begin{figure*}[]
	\subfigure[]{\includegraphics[angle=0,width=0.5\textwidth,clip=]{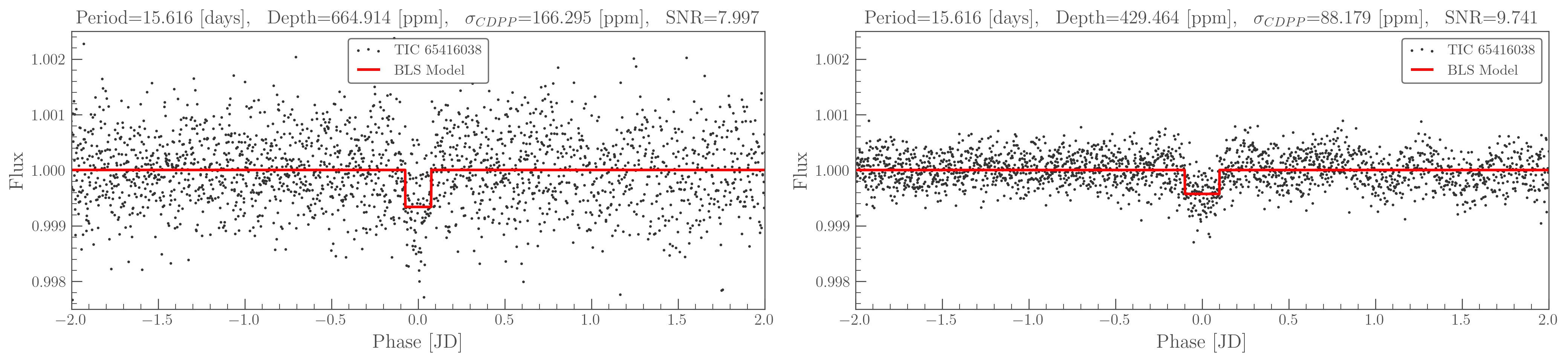}\label{fig:fig10}}		
	\subfigure[]{\includegraphics[angle=0,width=0.5\textwidth,clip=]{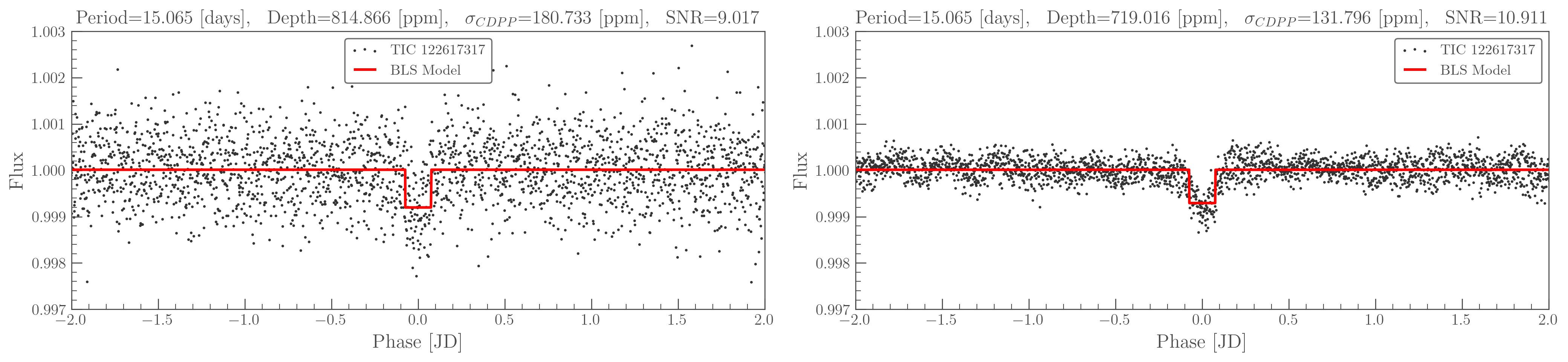}\label{fig:fig11}}		
	\subfigure[]{\includegraphics[angle=0,width=0.5\textwidth,clip=]{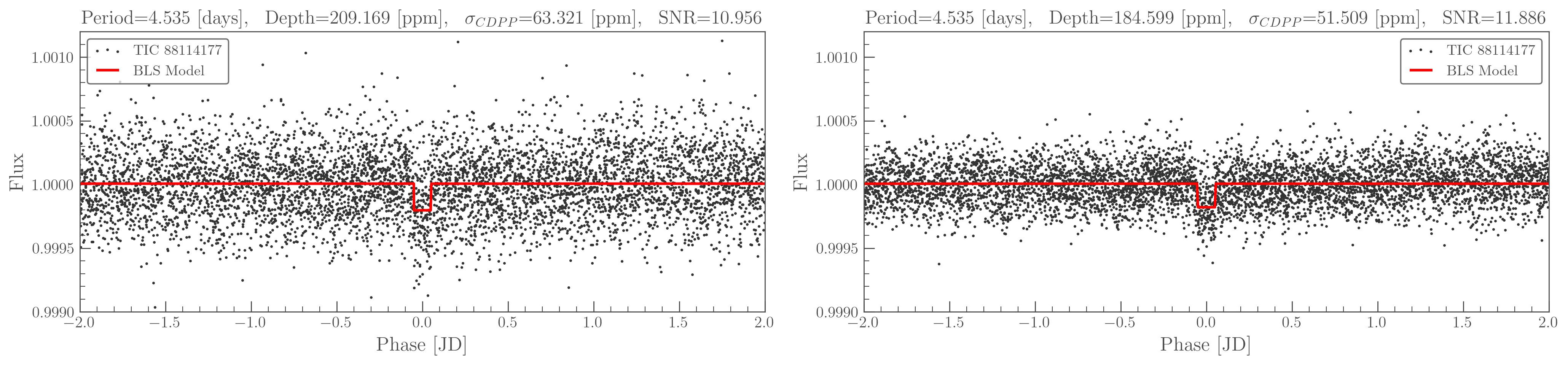}\label{fig:fig12}}	
	\subfigure[]{\includegraphics[angle=0,width=0.5\textwidth,clip=]{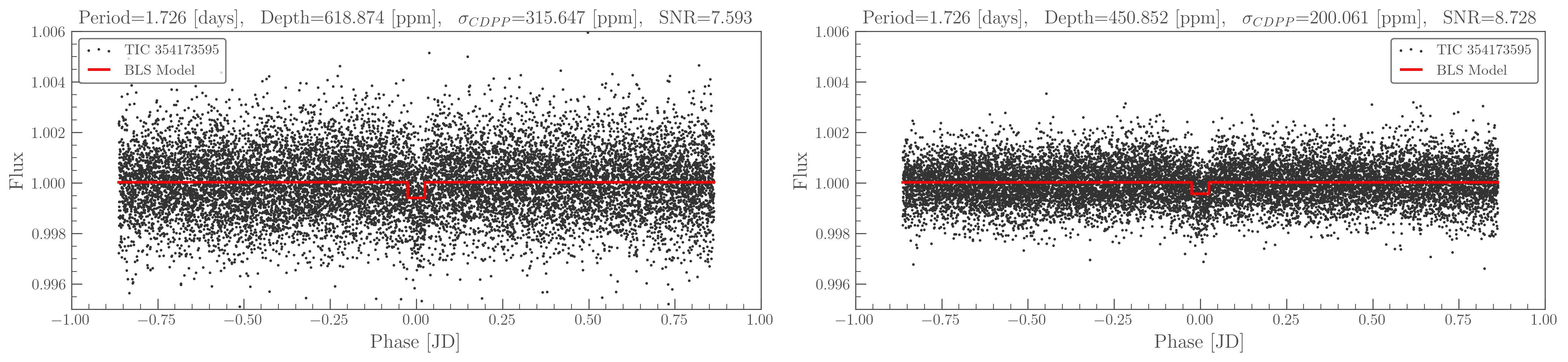}\label{fig:fig13}}	
	\subfigure[]{\includegraphics[angle=0,width=0.5\textwidth,clip=]{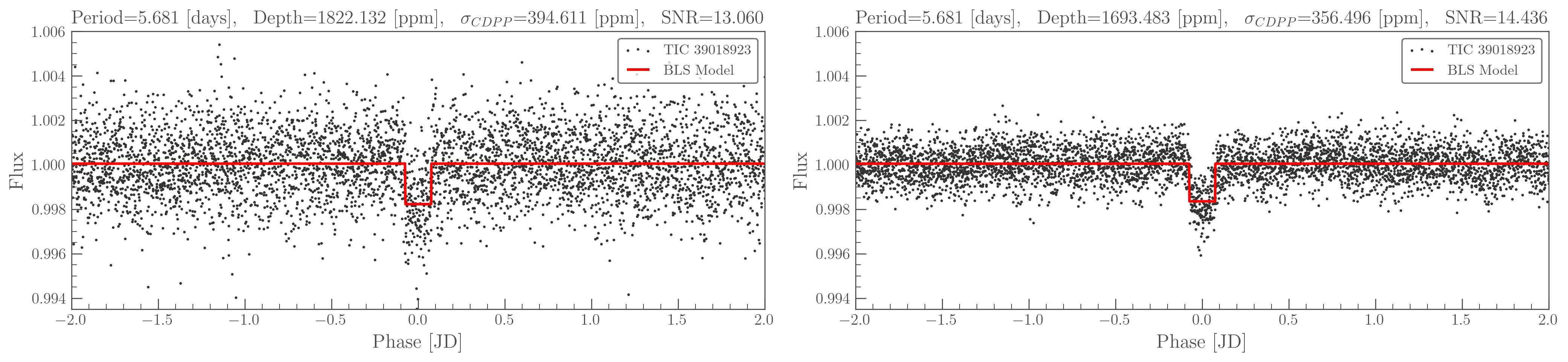}\label{fig:fig14}}		
	\subfigure[]{\includegraphics[angle=0,width=0.5\textwidth,clip=]{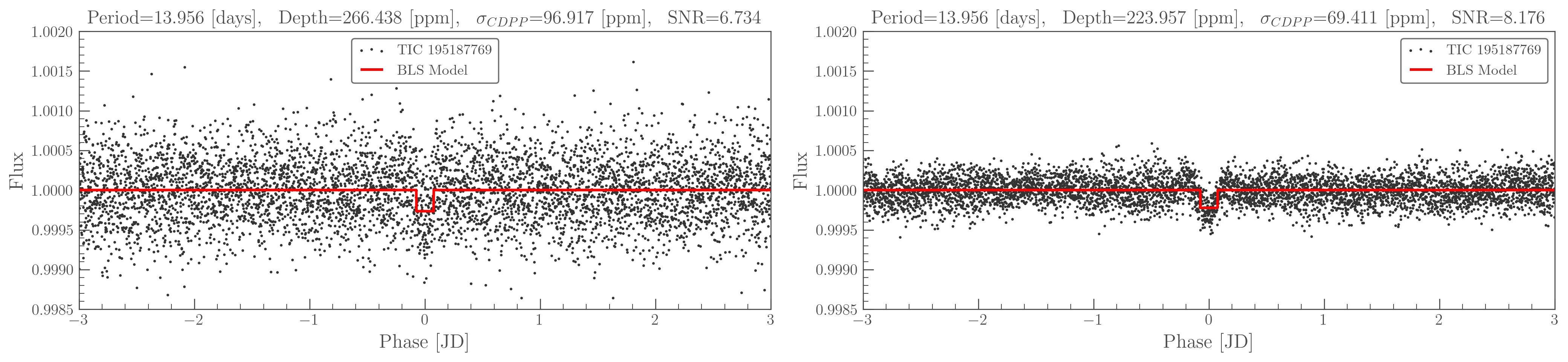}\label{fig:fig15}}		
	\caption{Folded light curves of six encouraging TESS candidates. The left side light curve in each panel corresponds to the initial time series, whereas the right side ones belong to the resulting reconstructed data from SSA. The red curve in each figure corresponds to the best transit model derived from the BLS algorithm. On top of each figure, Period, Depth, $\sigma_{cdpp}$ and SNR of the transits are reported. A decrease in fluctuations, improvements in the SNRs and changes in the best fit BLS models can be noted in the folded light curves after the SSA.  \label{fig:lc2}}
\end{figure*} 

\indent As mentioned in the introduction, the depth of an exoplanetary transit ($\delta$) is proportional to the ratio of squared planet to star radius \citep{2019Heller}. Consequently, if an exoplanet is relatively too small compared to its host star (as is the case with the terrestrial exoplanets), its transit signal would have a low SNR and may even remain unnoticeable in the fluctuations of the raw data (as demonstrated in the third simulation of section \ref{subsec: CS1}).

\noindent Additionally, the variability or the faintness of a host star or background and instrumental noises can cause a higher range of fluctuations in the light curve which would result in a higher $\sigma_{cdpp}$ and a lower SNR, respectively. As the SNR is also dependant on the number of transits (N in equation \ref{eq:SNR}), in exoplanetary transits with relatively long periods and the same amount of observations, the SNR would be lower as well.

\indent One of the main applications of SSA is time series de-noising and smoothing and it has been implemented as an effective noise removal technique in different contexts \citep{2014Zabalza, 2005feng}. As we have discussed in our case study in the previous section, SSA can also be used for de-noising a photometric time series in order to retrieve an exoplanetary transit signal. There are currently more than $250$ TESS project candidates in the Exoplanet Follow-up Observing Program\footnote{\url{https://exofop.ipac.caltech.edu/tess/view_toi.php}} (ExoFOP) \citep{2019Akeson} that show very low SNR transit-like signals.     

\noindent This indicates a great potential for SSA to investigate the true natures of these signals and provides a preliminary insight into whether there should be further photometric or spectroscopic observations. we implemented our SSA de-noising algorithm and compared the SNRs of these signals before and after this procedure. If the transit-like signal disappears (which manifests in a lower SNR) it can be deduced that the signal was due to minor background periodicities or a noise-like feature. Whereas, if the signal becomes more prominent (a higher SNR), the target star would make an appropriate candidate for follow-up photometric or spectroscopic observations to validate the transit. 

\noindent Furthermore, as low SNR transits have very weak signals that are infused with noise, the BLS model derived from the raw data may yield inaccurate planetary radii and other astrophysical parameters. In these situations, SSA can be used to resolve these issues.

\begin{deluxetable*}{ccccccccc}
\tablenum{1}
\tablecaption{Photometric accuracy and Transit data for the six TESS candidates displayed in figure \ref{fig:lc2}. By comparing columns corresponding to the initial and SSA time series, we can note the changes in transit depths, decreases in fluctuations ($\sigma_{cdpp}$) and improvements in SNRs. \label{tab:SNRs}}
\tablewidth{0pt}
\tablehead{\colhead{TIC ID}&\colhead{\rm{TOI}}&\colhead{\rm{P [days]}}&\colhead{$\delta$ [ppm] (initial)}&\colhead{$\sigma_{cdpp}$ [ppm] (initial)}&\colhead{SNR (initial)}&\colhead{$\delta$ [ppm] (SSA)}&\colhead{$\sigma_{cdpp}$ (SSA)}&\colhead{SNR (SSA)}}
\startdata
65416038 & 4190 & 15.616 & 664.914 & 166.295 & 7.996 & 429.464 & 88.179 & 9.741 \\
122617317 & 4311 & 15.065 & 814.866 & 180.733 & 9.017 & 719.016 & 131.796 & 10.911 \\
88114177 & 5531 & 4.535 & 209.169 & 63.321 & 10.956 & 184.599 & 51.509 & 11.886 \\
354173595 & 5974 & 1.726 & 618.874 & 315.647 & 7.593 & 450.852 & 200.061 & 8.728 \\
39018923 & 2216 & 5.681 & 1822.132 & 394.611 & 13.060 & 1693.483 & 356.496 & 14.436 \\
195187769 & 5550 & 13.956 & 266.438 & 96.917 & 6.734 & 223.957 & 69.411 & 8.176
\enddata
\end{deluxetable*}

\indent Figure \ref{fig:lc2} displays light curves of the most promising TESS candidates that showed an improved SNR after the SSA. The left side light curve in each panel corresponds to the initial time series, whereas the right side ones belong to the resulting reconstructed data. More information for each candidate (including the period, depth ($\delta$), averaged noise estimate ($\sigma_{cdpp}$) and SNR of the transit signal) is given in table \ref{tab:SNRs}.

\begin{figure}[]
\center
\includegraphics[angle=0,width=0.4\textwidth,clip=]{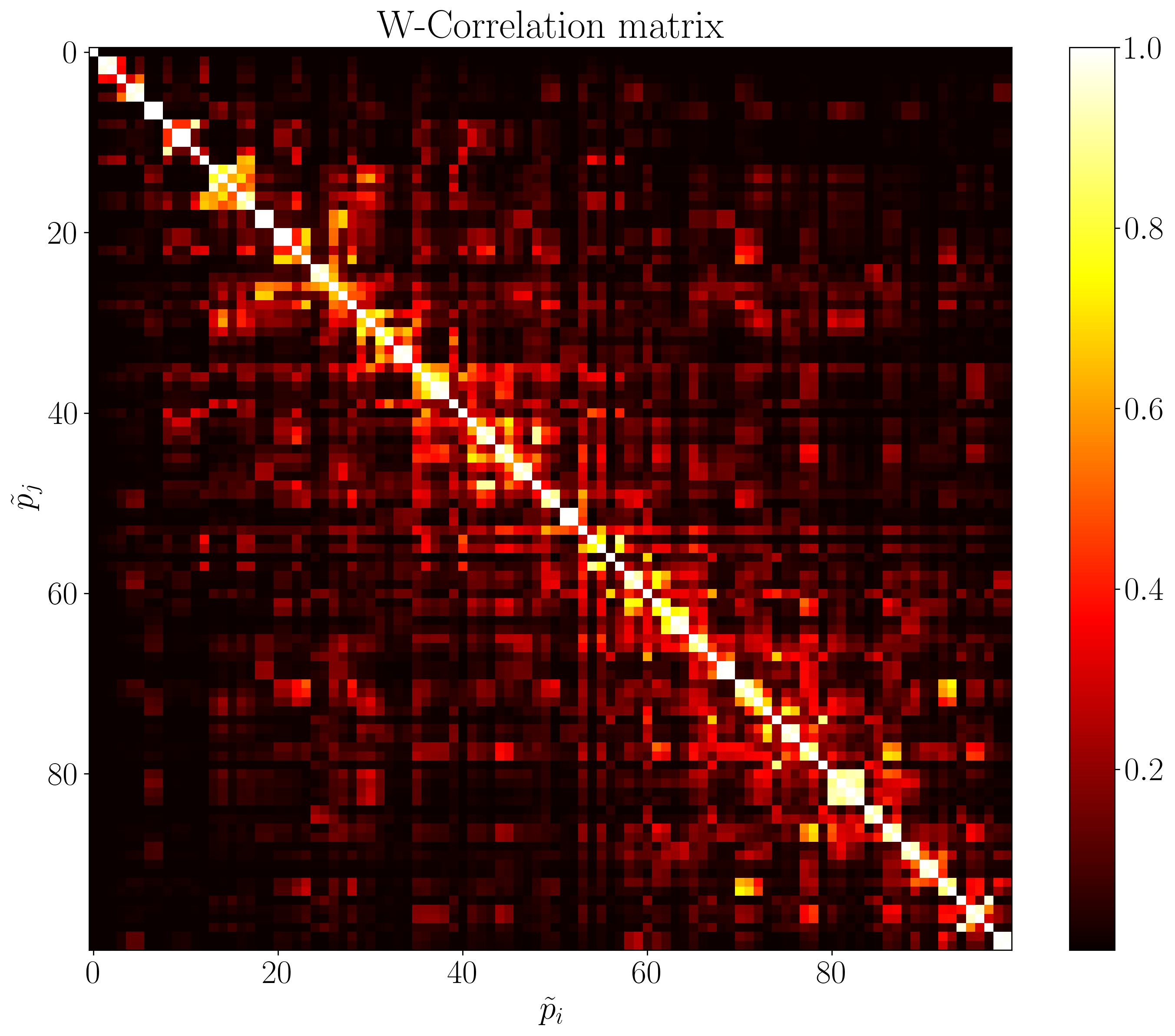}
\caption{Visualization of w-correlation matrix for TIC-$65416038$ with $L=100$.} \label{fig:DeW}
\end{figure}

\noindent As an example of how this process works, let us consider the folded time series of TIC-$65416038$ (figure \ref{fig:fig10}). There is a transit signal with period $P=15.616$ and $SNR=7.996$. The w-correlation matrix for a SSA decomposition of this time series is shown in figure \ref{fig:DeW}. After removing the components with a noise-like behaviour, the resulting reconstructed light curve is shown in the right side of figure \ref{fig:fig10}. The SNR of the transit signal has increased to $9.741$, which means that it is a genuine signal and not part of the underlying noise. Additionally, we can see differences in the durations and depths of the transit between the initial and the de-noised BLS models of the time series.

\indent In figure \ref{fig:lc2}, the SNRs of transit signals were improved by the SSA de-noising, but these improvements were less substantial compared to the simulations of case study 1 (section \ref{subsec: CS1}). This can be explained by considering the fact that as a result of longer transit periods and shorter observation, these TESS light curves have very small number of transits (N) in their available photometric time series. AS SSA increases the SNR by decreasing the noise-like fluctuations ($\sigma_{cdpp}$) in data, future observations of the same targets leads to a higher N and an increased SNR. Nevertheless, the planetary transits can be distinguished easier from the noise and other intrinsic fluctuations in the de-noised light curves. These candidates are very encouraging for further ground or space based observations.  

\section{Conclusions}
\indent Transit photometry is a very sensitive method for detecting and characterizing exoplanets. Photometric data of exoplanetary transits from space based missions like Kepler/k2 and TESS has considerably expanded our knowledge of planets beyond our Solar system and the accurate modeling and characterization of these transits is of great importance.

\indent Singular Spectrum Analysis is a powerful method for decomposing a time series into its constituting components. Reconstructing a new time series by selecting different groups of these components based on their weighted correlation values, can incorporate and separate the behaviour of a specific trend, periodic signal or different noise factors.    

\indent In this work, after discussing and reviewing the principles of basic SSA, we studied its further applications for the analysis of photometric time series by simulating different sets of time series incorporating planetary transit signals with low SNRs bellow the detection criteria ($SNR < 10$) . We have shown that by implementing SSA, we can separate the main variability trends of a star and the noise factors from its light curve and then use the BLS algorithm to look for transit-like signals with an increased SNR that could indicate the existence of a planetary object. Provided that the number of transits in a light curve is large enough (more than 5 transits per light curve) the increase in SNR can be as substantial as 30 units.

\noindent Additionally, In low SNR transit signals, the BLS algorithm may yield inaccurate results for the transit depth, duration or mid-transit time which in turn leads to the inaccurate inference of the astrophysical parameters of the planet. we have shown that SSA can be used as a noise removal technique in these circumstances for the accurate modeling of low SNR transiting exoplanets. 

\indent In this paper, we mainly applied our implementation of SSA to photometric time series from NASA's TESS mission, but this algorithm can be easily applied to other current or future ground and space based photometry surveys.

\indent A \textsc{Python} implementation of all the algorithms discussed in this paper can be found in \cite{hossein_fatheddin_2024_10938904}.

\section*{Acknowledgements}
This work was supported by the benefactors of the Swaantje Mondt fund. H. F. thanks Huub Rottgering, Wouter Schrier and Mozafar Allahyari for their help and support.

\indent This paper includes data collected by the TESS mission. Funding for the TESS mission is provided by the NASA's Science Mission Directorate. 

\indent This work has made use of data from the European Space Agency (ESA) mission {\it Gaia} (\url{https://www.cosmos.esa.int/gaia}), processed by the {\it Gaia} Data Processing and Analysis Consortium (DPAC, \url{https://www.cosmos.esa.int/web/gaia/dpac/consortium}). Funding for the DPAC has been provided by national institutions, in particular the institutions participating in the {\it Gaia} Multilateral Agreement. 

\indent This research made use of \textsc{LightKurve}, a \textsc{Python} package for Kepler and TESS data analysis \citep{2018Lightkurve}.

\indent We thank the reviewer for useful comments and suggestions.

\newpage
\bibliography{v2}{}
\bibliographystyle{aasjournal}
\end{document}